\documentclass[aps,prl,reprint,groupedaddress,showpacs]{revtex4-1}

\usepackage{graphics,color,epsfig}
\usepackage{bbm}
\usepackage{amsmath,amsfonts,amssymb,amsthm}
\newcommand\E{{\mathbb E}}
\newcommand\cR{{\mathcal R}}

\begin{document}
\title{Achlioptas processes are not always self-averaging}

\author{Oliver Riordan}
\altaffiliation{These authors contributed equally to this work.}
\affiliation{Mathematical Institute, University of Oxford, 24--29 St Giles', Oxford OX1 3LB, United Kingdom}

\author{Lutz Warnke}
\altaffiliation{These authors contributed equally to this work.}
\affiliation{Mathematical Institute, University of Oxford, 24--29 St Giles', Oxford OX1 3LB, United Kingdom}

\date{May 24, 2012}

\begin{abstract}
We consider a class of percolation models, called Achlioptas processes, 
discussed in [Science {\bf 323}, 1453 (2009)] and [Science {\bf 333}, 322 (2011)]. 
For these the evolution of the 
order parameter (the rescaled size of the largest connected component) has been 
the main focus of research in recent years. 
We show that, in striking contrast to `classical' models, self-averaging is not 
a universal feature of these new percolation models: there are natural 
Achlioptas processes whose order parameter has random fluctuations that do not 
disappear in the thermodynamic limit. 
\end{abstract}

\pacs{64.60.ah, 05.40.-a, 02.50.Ey, 89.75.Hc, 64.60.aq}

\maketitle

\section{Introduction}\label{sec:intro}
Percolation is a fundamental problem in statistical physics, and the emergence 
of long range connectivity in various percolation models is one of the 
quintessential examples of a phase transition. 
While many `classical' models and their broad universality classes are 
nowadays well understood~\cite{ASPerc,GPerc,BRPerc}, in recent years a new 
class of percolation models has been widely studied. 
These so-called \emph{Achlioptas processes}~\cite{Science2009,TBDK2006,JSNW2007,AAP2011} 
are defined via a slight modification of well-studied Erd\H{o}s--R\'enyi 
random graphs, and have been of great interest to many physicists~\cite{Science2009,
PhysRevLett.103.045701,
PhysRevLett.103.135702,
PhysRevLett.103.168701,
PhysRevLett.103.255701,
PhysRevE.81.036110,
PhysRevLett.104.195702,
PhysRevE.82.042102,
PhysRevE.82.051105,
dCDGM2010,
Manna2011177,
Nature2011,
PhysRevLett.106.095703,
PhysRevLett.106.115701,
PhysRevE.83.032101,
PhysRevE.83.031133,
PhysRevLett.106.225701,
Science2011,
PhysRevE.84.020101,
PhysRevE.84.020102,
PhysRevE.84.066112,
PhysRevLett.107.275703,
Tian2012286,
SS20122833} due to their intriguingly different features.

One of the most studied properties of Achlioptas processes is the 
evolution of the order parameter (the rescaled size of the largest connected component), 
yielding several surprises~\cite{Science2009,dCDGM2010,AAP2011,Science2011}. 
Indeed, certain Achlioptas processes were first claimed to have 
discontinuous (first order) phase transitions~\cite{Science2009}, in striking 
contrast to the typical second order transition observed in classical 
percolation models. Called \emph{explosive percolation}, this was subsequently 
supported by many researchers~\cite{PhysRevLett.103.045701,PhysRevLett.103.135702,PhysRevLett.103.168701,PhysRevLett.103.255701,PhysRevLett.104.195702,PhysRevE.81.036110,PhysRevE.82.042102,PhysRevE.82.051105,PhysRevE.83.032101}, 
but recently it was mathematically rigorously proven that the  
transition is in fact continuous for \emph{all} mean-field Achlioptas 
processes~\cite{AAP2011,Science2011}, although it can be extremely 
steep~\cite{dCDGM2010}. Furthermore, compared to the classical cases many of these 
models seem to behave differently in essential ways, representing a new universality 
class~\cite{dCDGM2010,PhysRevLett.106.225701,PhysRevE.84.066112,Tian2012286} 
whose basic features still require further 
investigation~\cite{BohmanScience,JansonScience}.

In this letter we show that these new percolation models can show another 
surprising behaviour: there are natural Achlioptas processes whose order  
parameter has large random fluctuations, i.e., is not self-averaging. 
This is in contrast to classical models, where the order parameter 
converges to a nonrandom function in the thermodynamic limit. 
As our nonconvergent examples are from different universality 
classes, including the `explosive' one, these also serve as a cautionary 
tale: when studying a wide range of Achlioptas processes, one should not 
take convergence for granted.

\section{The Model}\label{sec:model}
Many `competitive' percolation models on the complete graph have been studied, 
see e.g.~\cite{Science2009,TBDK2006,JSNW2007,dCDGM2010,PhysRevLett.104.195702,Nature2011,AAP2011} 
and the references therein. 
These Achlioptas processes start, as in the classical 
Erd\H{o}s--R\'enyi (ER) model, with an empty graph consisting of a large number $n$ 
of isolated vertices, and then sequentially add random edges to the graph. 
In the simplest case in every round two random edges $e_1=v_1v_2$ and $e_2=v_3v_4$ 
are picked (rather than one), and, using some rule, one of them 
is chosen and added to the evolving graph. Of course, there is nothing special 
about two edges; indeed, Achlioptas processes are a subset of the general 
class of \emph{$\ell$-vertex rules} introduced in~\cite{AAP2011}. 
Here in every round $\ell$ uniformly random vertices $v_1, \ldots, v_{\ell}$ 
are chosen, and two of them are connected with an edge using some rule 
(see~\cite{AAP2011} for more variations). 
Note that the resulting percolation processes include the ER model, which we 
obtain by always adding~$v_1v_2$.

In the following we introduce two different $3$-vertex rules belonging to 
different universality classes, always writing $c_i$ for the size of the 
component containing $v_i$. 
The first rule, the \emph{NG rule}, proceeds as follows. If all three component 
sizes $c_i$ are equal, add $v_1v_2$. If exactly two component sizes $c_i$ are 
equal, connect the  corresponding vertices with an edge. Otherwise (if all $c_i$ 
are different) join the vertices in the two smallest components. This rule is 
named after Nagler and Gutch, who suggested a slight variant in a different 
context (personal communication). It is a  modification of the `explosive' 
\emph{triangle rule} introduced in~\cite{PhysRevLett.103.255701,PhysRevLett.104.195702}, 
which has a steep (but continuous) transition. The rapid growth of the  
order parameter in Figure~\ref{fig:L1} shows that NG  also  belongs to the 
class of `explosive' rules.

The second rule we consider can be viewed as a modified ER model and thus we
use the shorthand \emph{mER}. Writing $L_1$ and $L_2$ for the sizes of the 
two largest components of the evolving graph, it is defined as follows. 
If the two largest components in the current graph have the same size 
($L_1=L_2$), add $v_1v_2$. When $L_1>L_2$, if at least two $c_i$ are equal to 
$L_1$, connect two corresponding vertices; otherwise connect two vertices in 
components of size smaller than $L_1$. Figure~\ref{fig:L1} indicates that mER 
belongs to the class of `nonexplosive' rules, where the order parameter evolves 
rather slowly around the percolation threshold (as in the classical ER case).

\section{Results}\label{sec:exp} 
For Achlioptas processes the natural order parameter is 
\[ \rho_n(t)=\frac{L_1(tn)}{n}, \] 
where $L_1(m)$ denotes the size of the 
largest connected component after $m$ steps (suppressing in the notation the 
dependence on the rule $\cR$ and on the number of vertices $n$). 
For classical percolation models it is well known that the order parameter 
converges to a nonrandom function in the thermodynamic limit, i.e., that there 
exists a \emph{scaling limit} $\rho:[0,\infty)\to[0,1]$ such that 
\begin{equation}\label{SL}
\lim_{n \to \infty} \rho_n(t) = \rho(t).
\end{equation}
Since $L_1(tn)$ is random, this actually means that 
$\rho_n(t)=L_1(tn)/n$ converges in probability to $\rho(t)$, see 
e.g.~\cite{AAP2011} for a more detailed discussion. 
So \eqref{SL} asserts: 
(A) that $\rho_n(t)$ is self-averaging, i.e., that the random fluctuations are 
negligible as $n \to \infty$, 
and 
(B) that the expected value of $\rho_n(t)$ converges to some deterministic
function, i.e., that $\E\rho_n(t)$ does not depend on $n$ in the 
thermodynamic limit.

\begin{figure}[t]%
\hspace*{-1.0em}\epsfig{file=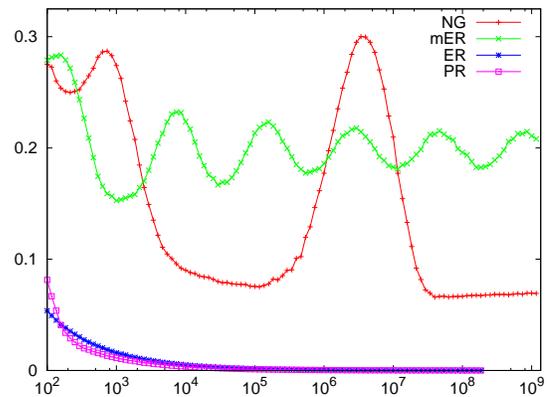,width=3.1in}\\%
\caption{(Color online). 
Estimated value of $\sigma_n/\mu_n$ for $\rho_n(1)$ as a function of $n$ (each 
based on $10^4$ runs) using the NG, mER, ER rules and product rule (PR)~\cite{Science2009}. 
Self-averaging (ER, PR) does not correlate with being `explosive' (NG, PR).\vspace*{-0.05em}\label{fig:SDAVG}}%
\end{figure}

\begin{figure}[t]%
\hspace*{-1.0em}\epsfig{file=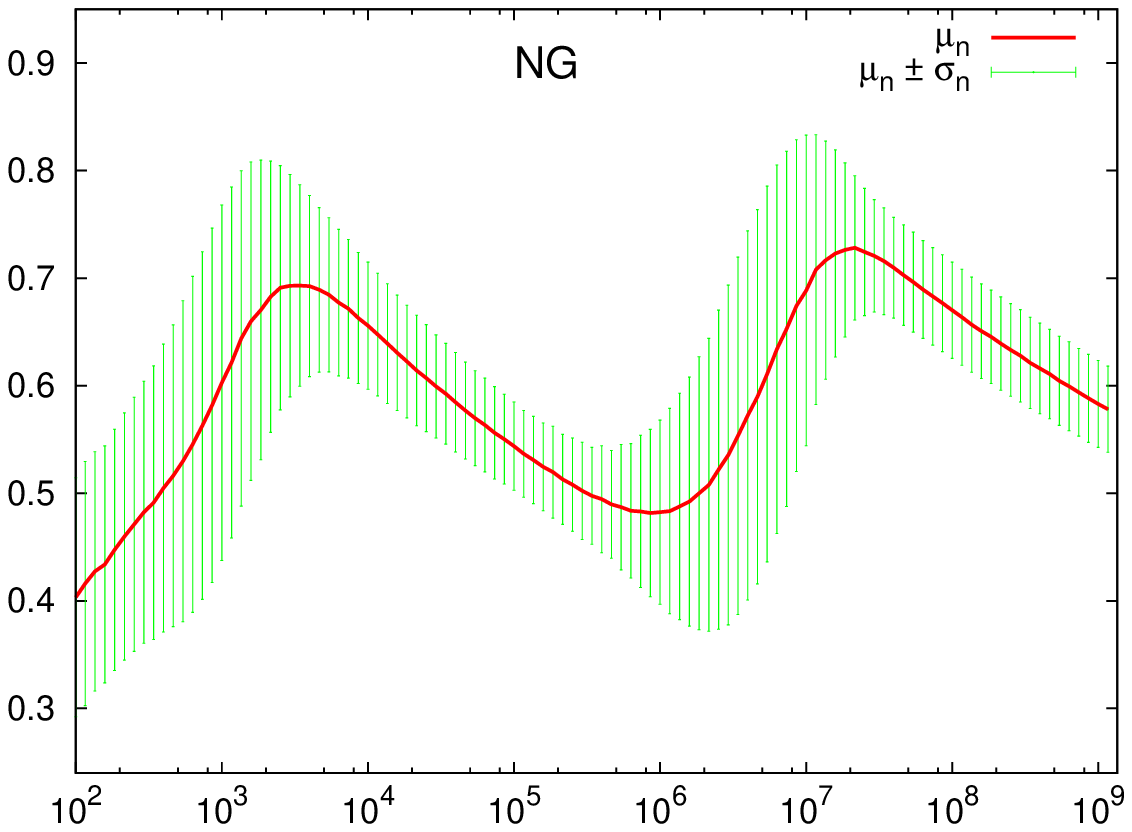,width=3.1in}\\%
\hspace*{-1.0em}\epsfig{file=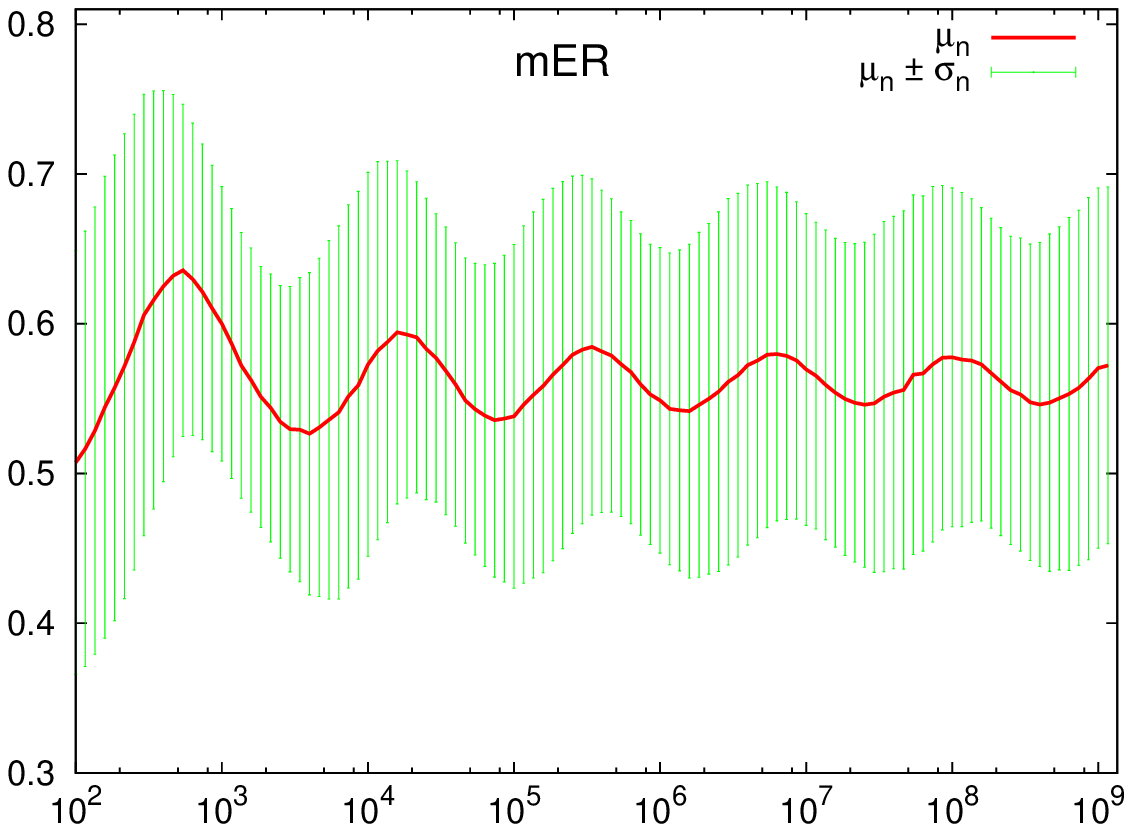,width=3.1in}\\%
\caption{(Color online). Simulation of the order parameter $\rho_n(1)=L_1(n)/n$ 
as a function of $n$ for the NG and mER rules: average and 
standard deviation ($10^4$ runs for each $n$).\vspace*{-0.05em}\label{fig:L1:mean}}%
\end{figure}

In the following we show that for the NG and mER rules the corresponding 
scaling limits do \emph{not} exist, i.e., that \eqref{SL} fails. 
To establish this we estimate the mean $\mu_n$ and standard 
deviation $\sigma_n$ of $\rho_n(1)$ for a range of $n$ (based on $10^4$ 
runs for each $n$). Formally we say that $\rho_n(t)$ is \emph{self-averaging} 
(at time $t=1$) if $\sigma_n/\mu_n \to 0$ as $n \to \infty$, 
which is well known to hold for the ER process (see also Figure~\ref{fig:SDAVG}). 
For the NG rule, and especially the mER rule, Figure~\ref{fig:SDAVG} 
demonstrates that $\sigma_n/\mu_n$ does \emph{not} tend to $0$ in 
the thermodynamic limit; hence these are \emph{not} self-averaging. 
This is further illustrated in Figure~\ref{fig:L1:mean}, which also indicates 
that the expected values of the corresponding order parameters have a 
non-trivial dependence on $n$: they seem to oscillate. 
While for the mER rule the amplitudes seem to vanish in the thermodynamic 
limit, for the NG rule we leave it open whether $\E\rho^{\mathrm{NG}}_n(t)$ 
converges as $n \to \infty$ or not: if it does, then Figure~\ref{fig:L1:mean} 
suggests that this only happens for extremely large values of $n$.

To put our findings into perspective, note that it is easy to construct  
artificial Achlioptas processes without scaling limits (using, for example, 
different rules  depending on the vertices sampled in the first round). 
The point is that the mER, and in particular the NG rule, are natural examples 
which nevertheless are nonconvergent. Furthermore, they belong to two very 
different classes of rules (explosive and nonexplosive). 
This shows that, contrary to classical percolation models, 
self-averaging is \emph{not} a universal feature of Achlioptas processes.

\section{Analysis}\label{sec:analyis} 
We start by collecting some structural properties of Achlioptas processes. 
Remark~9 in~\cite{AAP2011} implies that whp (meaning with high probability, 
i.e., with probability tending to $1$ as $n \to \infty$) the following two 
properties hold throughout the entire evolution of any $\ell$-vertex Achlioptas 
process: 
(i)~the vertex set of the union of the $\ell-1$ largest components changes 
by at most $o(n)$ vertices in any $o(n)$ steps, 
and (ii)~there exists a function $s=s(n)$ such that the $\ell$-th largest 
component always has size at most $s=o(n)$. 
Although (i) might seem rather technical at first sight, it has an important 
implication: it shows that (whp) the size of the largest component can `jump', 
i.e., change by $\Theta(n)$ vertices in $o(n)$ steps, only if there is a step 
where two linear size components merge to form the new largest component (for 
$\ell$-vertex rules this strengthens some of the main conclusions 
in~\cite{Nature2011}).

\begin{figure}[t]%
\hspace*{-1.0em}\epsfig{file=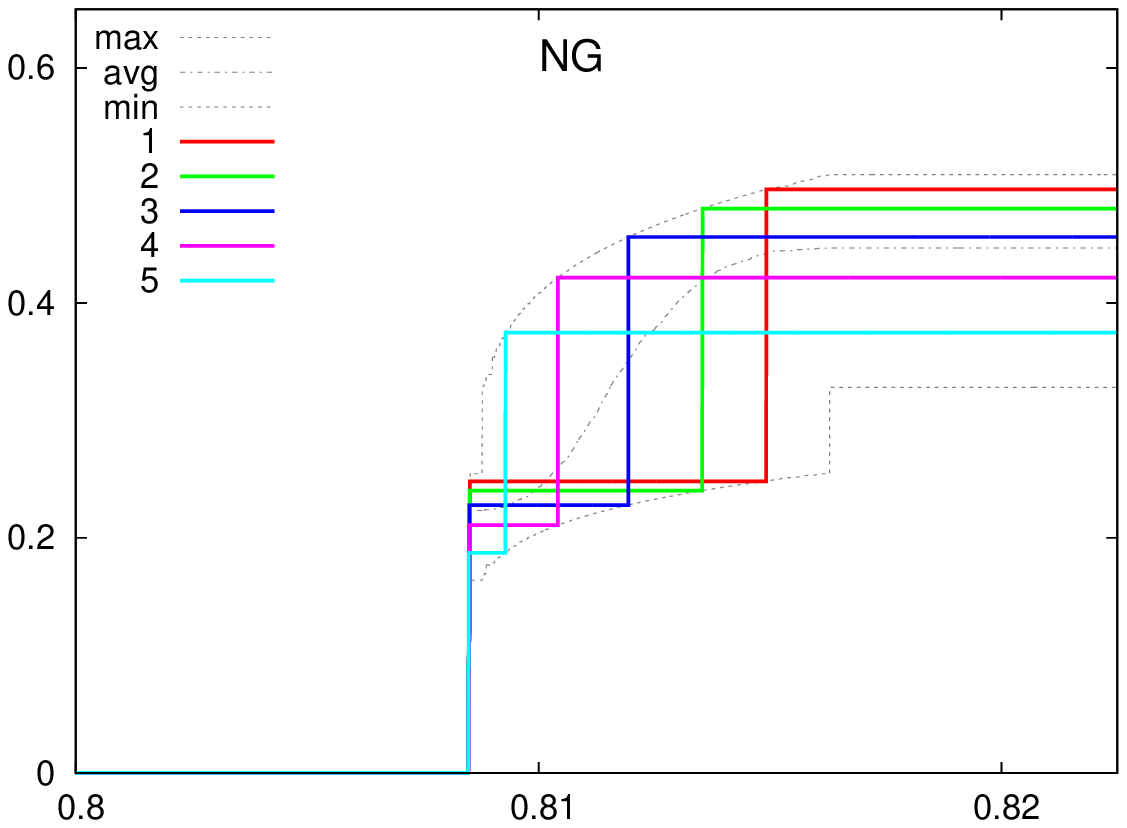,width=3.1in}\\%
\hspace*{-1.0em}\epsfig{file=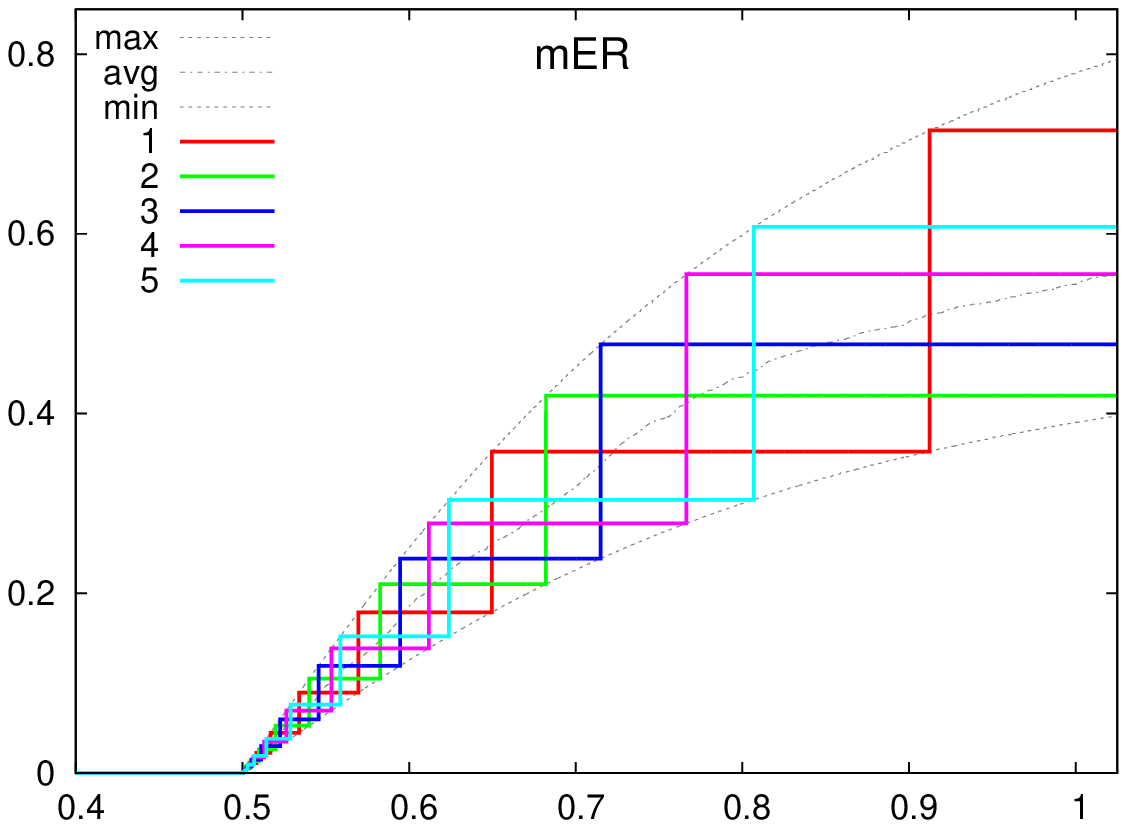,width=3.1in}%
\caption{(Color online). Simulation of the order parameter $\rho_n(t)=L_1(tn)/n$ 
as a function of $t$ for the NG and mER rules with $n=10^{11}$: five sample 
runs together with min, max and average of $10^3$ runs.\label{fig:L1}}%
\end{figure}%

Turning to the evolution of the NG and mER rules, the key property of both is 
that they prevent the largest component from growing as long as the two largest 
components have different sizes ($L_1 > L_2$). But, whenever we have two 
linear size `giant' components of the same size, they merge with constant 
probability in each step (recall that by property~(ii) we can have at most two 
such components as here $\ell=3$).  
If this happens then $L_1$ doubles and by~(ii) we are left with one giant; all 
other components, in particular $L_2$, have size at most $s=o(n)$. Now $L_1$ is  
prevented from growing and $L_2$ slowly starts growing. After some time $L_2$ 
emerges to linear size, while all smaller components have size at most $s=o(n)$ 
by~(ii). When $L_1$ and $L_2$ have similar size, they might overtake each other 
repeatedly, but since we have many small components they will merge in $o(n)$ 
steps, and again $L_1$ essentially doubles by property~(i). 
To summarize, after the first linear size giant appears, the remaining 
evolution of $L_1$ is essentially governed by discrete doublings.

The behaviour at the percolation threshold, where the first linear size 
component is formed, seems rather complicated for both rules. However, the 
simulation results for the NG and mER rules depicted in Figure~\ref{fig:L1} 
 show that the size of the linear size giant 
component is \emph{not} deterministic once it first emerges: note the
large spread in the order parameter $\rho_n(t)=L_1(tn)/n$. 
So, after the percolation threshold the largest component has random size. 
Since the remaining discrete doublings depend on this initial value, this 
shows that the corresponding $\rho^{\mathrm{NG}}_n(t)$ and $\rho^{\mathrm{mER}}_n(t)$ 
exhibit large random fluctuations in their subsequent evolution. 
Figures~\ref{fig:SDAVG} and~\ref{fig:L1:mean} indicate that these do not 
disappear in the thermodynamic limit: the order parameters of the NG and mER 
rules are not self-averaging. 
This `freezing in' of early variations (the microscopic fluctuations in the 
size of the largest component are magnified and propagated to later in the 
process) is similar to the nonconvergent behaviour of the maximum degree in 
the Barab\'asi--Albert network model~\cite{Science1999}, see e.g.~\cite{BR-SF}.

For the mER rule we can formally describe this propagation to later stages, 
since its evolution has a close connection to Erd\H{o}s--R\'enyi 
random graphs. To see this, we first argue that throughout its evolution, up 
to $o(n)$ differences, mER yields a graph of the following form: its largest 
component has size $\alpha n$, where $\alpha \ge 0$, and the remaining smaller 
component sizes follow the distribution of an ER graph 
$G(\tilde{n},\lambda/\tilde{n})$ where $\tilde{n}=(1-\alpha)n$ and 
$\lambda \ge 0$; here $G(n,p)$ denotes the \emph{binomial random graph} model 
in which, starting with an empty graph on $n$ vertices, each of the 
$\binom{n}{2}$ possible edges is included independently with probability $p$ 
(for $p= 2m/n^2$ it is well known to be  very similar to the uniform ER graph 
obtained after inserting $m$ random edges). 
The key observation is that whenever $L_1 > L_2$, the mER rule either 
adds an edge with both ends inside the largest component, or one with both 
ends outside. In each case the endpoints are chosen uniformly at random. 
So, between each doubling of $L_1$ the 
graph outside the largest component evolves like a (rescaled) ER graph. 
Once $L_1$ and $L_2$ have similar linear size they merge in $o(n)$ steps, 
so, for equality up to $o(n)$ differences, it suffices to show that the step in 
which they merge retains the claimed form of the graph. 
But this follows from the \emph{discrete duality principle} for 
$G(n,c/n)$ first used in~\cite{BB-evo}: the removal of the largest component 
leaves an ER graph $G(n',d/n')$ on $n'$ vertices with $d \le 1$. 
Summarizing, the evolution of the mER graph after $m=tn$ steps is described 
by the evolution of the parameters $\alpha(t)$ and $\lambda(t)$, which control 
the size of the largest component and the distribution of the smaller component 
sizes, respectively. 
For $t \le 1/2$ we have $\alpha=0$ and $\lambda=2t$: the graph is 
still close to the ER graph (as all components are small). Close to $t=1/2$ the 
parameters evolve in a non-deterministic way. Then, as a function of this 
`initial randomness' at time $t_0>1/2$, the later evolution is 
deterministic and can be described explicitly for any $t \ge t_0$. 
For the mER rule this explains how the lack of self-averaging depicted in 
Figure~\ref{fig:L1} arises.

Finally, the `jumps' of the order parameter in Figure~\ref{fig:L1} show that 
the NG and mER rules are examples of Achlioptas processes which violate 
($\star$)~continuity throughout the process, although these satisfy the 
widely studied ($\dagger$)~continuity at the percolation threshold 
by~\cite{AAP2011,Science2011}. In fact, since in both rules the largest 
component grows essentially only by discrete doublings, it follows that after 
the percolation threshold both give rise to an `infinite' number of 
discontinuities in sense~($\star$). Property~(i) implies that in $\ell$-vertex 
rules such discontinuities can only arise if two linear size components merge. 
It would be interesting to know if, when restricting to $\ell$-vertex rules 
whose decisions (which edge to add) depend only on the component sizes 
$(c_1, \ldots, c_{\ell})$, this is the only mechanism leading to  
nonconvergent behaviour.

\section{Conclusions}\label{sec:concl} 
In summary, we have shown that self-averaging of the order parameter is 
\emph{not} a universal feature of Achlioptas processes: the corresponding 
scaling limits do not always exist. 
While for any particular rule this can be tested numerically via simulations 
(as in~\cite{PhysRevLett.106.225701,PhysRevE.84.020101} for example), the 
large interest in such percolation models~\cite{Science2009,
PhysRevLett.103.045701,
PhysRevLett.103.135702,
PhysRevLett.103.168701,
PhysRevLett.103.255701,
PhysRevE.81.036110,
PhysRevLett.104.195702,
PhysRevE.82.042102,
PhysRevE.82.051105,
dCDGM2010,
Manna2011177,
Nature2011,
PhysRevLett.106.095703,
PhysRevLett.106.115701,
PhysRevE.83.032101,
PhysRevE.83.031133,
PhysRevLett.106.225701,
Science2011,
PhysRevE.84.020101,
PhysRevE.84.020102,
PhysRevE.84.066112,
PhysRevLett.107.275703,
Tian2012286,
SS20122833} 
indicates that a theoretical investigation of convergence in Achlioptas 
processes is needed, see also~\cite{BohmanScience,JansonScience}. 
Is there a general criterion which allows us to determine whether certain rules 
are convergent or not?
Recently self-averaging has been rigorously established for a restricted class 
of Achlioptas rules~\cite{Unique2011}: this includes the `dCDGM' rule 
considered in~\cite{dCDGM2010}, but not the widely studied~\cite{Science2009, 
PhysRevLett.103.045701,
PhysRevLett.103.135702,
PhysRevLett.103.168701,
PhysRevLett.103.255701,
PhysRevE.81.036110,
PhysRevE.82.042102,
PhysRevE.82.051105,
Manna2011177,
Nature2011,
PhysRevLett.106.095703,
PhysRevE.83.032101,
PhysRevE.83.031133,
PhysRevLett.106.225701,
Science2011,
PhysRevE.84.020101,
PhysRevE.84.020102,
PhysRevE.84.066112,
PhysRevLett.107.275703,
Tian2012286,
SS20122833} product rule (PR).  
Simulations indicate that the scaling limit of the PR rule most likely exists 
(see e.g.\ Figure~\ref{fig:SDAVG}), but given the surprises that it has shown 
so far~\cite{Science2009,BohmanScience,Science2011,JansonScience} we can only 
be sure once we have rigorous results.

\vspace{-0.45em}

\end{document}